\documentclass[conference]{IEEEtran}
\IEEEoverridecommandlockouts
\usepackage{cite}
\usepackage{amsmath,amssymb,amsfonts}
\usepackage{algorithmic}
\usepackage{graphicx}
\usepackage{textcomp}
\usepackage{xcolor}
\usepackage{subcaption}
\usepackage{lipsum}
\usepackage{mathtools}
\usepackage{multirow}

\usepackage{cuted}
\def\BibTeX{{\rm B\kern-.05em{\sc i\kern-.025em b}\kern-.08em
    T\kern-.1667em\lower.7ex\hbox{E}\kern-.125emX}}
\begin{document}

\title{Normalized Benchmarking of Hybrid Switched-Capacitor DC-DC Converters}

\author{\IEEEauthorblockN{Gaël Pillonnet}
\IEEEauthorblockA{\textit{University of Grenoble Alpes} \\
\textit{CEA-LETI}\\
Grenoble, 38000, France \\
gael.pillonnet@cea.fr }
\and
\IEEEauthorblockN{Mahmoud Hmada, Patrick P. Mercier }
\IEEEauthorblockA{\textit{Department of Electrical \& Computer Engineering} \\
\textit{University of California San Diego}\\
La Jolla, CA 92037, USA \\
p.mercier@ucsd.edu}

}

\maketitle

\begin{abstract}
Hybrid switched-capacitor converters (HSCC) offer great potential for high efficiency and power density compared to purely capacitor- or inductor-based converters. However, the recent proliferation of HSCC topologies has made it difficult to choose the best one for a particular application. This paper presents a benchmarking framework that allows for direct comparison of popular HSCC topologies by analyzing various performance metrics such as passive component volume and bandwidth. By comparing all topologies at the same efficiency, same inductor ripple, and same output voltage ripple, this approach generates guidelines for topology selection and optimization, which can aid in wider industrial adoption and exploration of new topologies.
\end{abstract}


\section{Introduction}
\textcolor{black}{Leveraging} the performance of inductive-based converters with an additional first-stage switched capacitors network has been well-known in the literature \cite{middlebrook_transformerless_1988}, but has been revived by recent developments in applications such as new USB standards, high power computing power delivery, and 48V bus level adoption. Consequently, hybrid switched-capacitor converters (HSCC) (Fig. 1) have received particular attention in the last decade to address non-isolated high voltage conversion ratios (VCR) \cite{nishijima_analysis_2005,yousefzadeh_three-level_2006,pilawa-podgurski_merged_2008}. Compared to pure switched-capacitor converters (SCC), many (though not all) HSCC offer soft-charging operation, which eliminates the main SCC bottlenecks: charge-sharing losses \cite{lei_soft-charging_2014} and non-lossless regulation capability \cite{stauth_pathways_2018}. Compared to pure inductive-based converters, additional flying capacitors $C_F$ block a portion of the input voltage, allowing the use of low-voltage rating switches and enabling higher switching frequency operation \cite{kesarwani_resonant_2015}. The reasons for HSCC's supremacy lie in technological limitations on passive components, namely the relatively low energy density of inductors compared to capacitors \cite{perreault_opportunities_2009}, and on active devices, namely the negative impact of the blocking voltage on switch performance \cite{baliga_power_1989, stauth_pathways_2018, liu_97_2020}.

The choice of HSCC topology is crucial as it distributes the constraints over the three main converter components (switches, capacitors, and inductors). Influenced by the formalization of SCCs in the 2000s, topology zoology is mainly derived from SCCs: Dickson, series-parallel, Fibonacci, etc. The combination of a traditional SCC with inductors at the output provides a plethora of possible topological solutions, necessitating a multifaceted comparison between them. Previous contributions have already proposed some powerful comparison tools for predicting the minimal achievable output resistance \cite{pasternak_equivalent_2016, ye_modeling_2022, liu_97_2020}, mostly inspired by modeling that is dedicated to SCCs \cite{seeman_analysis_2008, le_design_2011}. In this paper, we propose to augment the effort in \cite{lei_analytical_2015} by introducing an improving framework for a normalized benchmark of fully soft-charging non-resonant HSCC topologies, referred to the baseline 2-level buck \textit{at the same power efficiency, and same inductor current and output voltage ripples}.

\begin{figure}[t]
    \begin{center}
    \captionsetup{justification=centering, belowskip=-20pt}
    \includegraphics{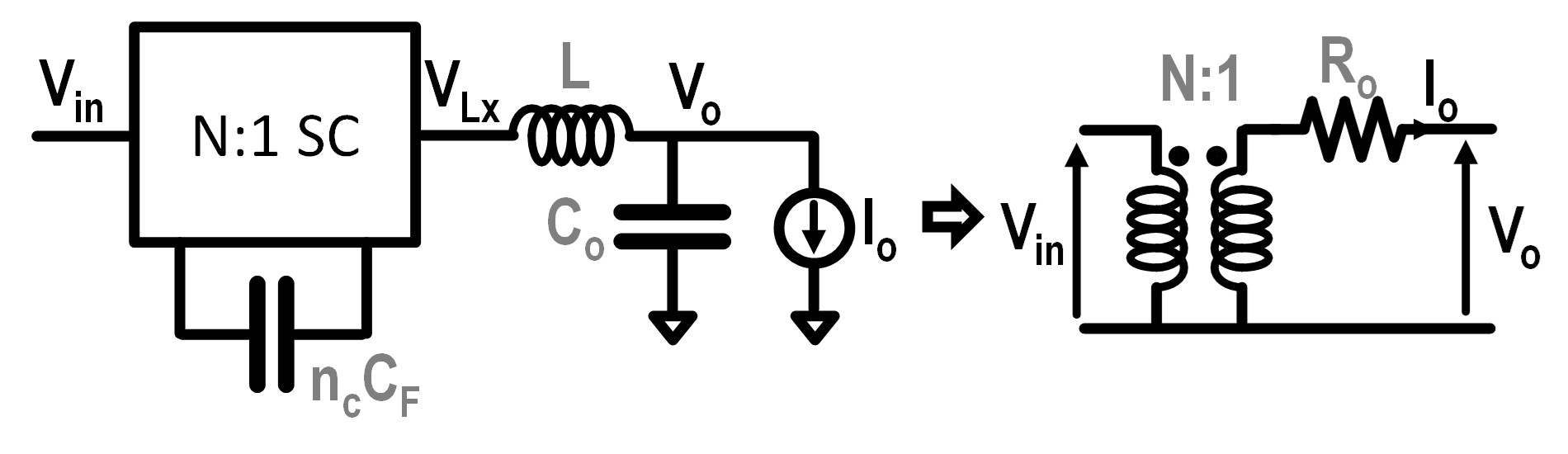}
    \caption{Generalized hybrid switched-capacitor converter structure and behavioral model.}
    \label{fig:1}
    \end{center}
\end{figure}

Our approach differs from previous works as we establish a connection between the intrinsic performance of passive and active component utilization through switch area and switching frequency amongst different topologies. This enables a comparison of all topologies under the same conduction and switching losses, including inductor losses, while maintaining the same inductor current and output voltage ripples for a fair comparison. Additionally, our analysis is conducted on a regulated HSCC, where the "output buck converter" achieves voltage regulation by adjusting the duty cycle. In other words, the VCR $M$ is not necessarily the inherent unregulated DC-DC transformation ($N$:1) given by the capacitive network. Our comparison is purely dimensionless, meaning it remains true for any input/output voltage/current levels, which helps to make the conclusions relatively general. Although this work provides a comparison in a particular set of input parameters, the method can be numerically solved under various constraints (switch scaling law, VCR) to extend the benchmark to a specific targeted design space.

The objective of this paper is more to point out the pros and cons of each topology than to reveal the best-in-class HSCC. The framework provides a direct inductor volume, bandwidth, voltage rating comparison among different previously introduced HSCC, letting designers to choose the best compromise for their targeted applications.

\begin{figure*}
    \begin{center}
    \captionsetup{justification=centering, belowskip=-20pt}
    \includegraphics{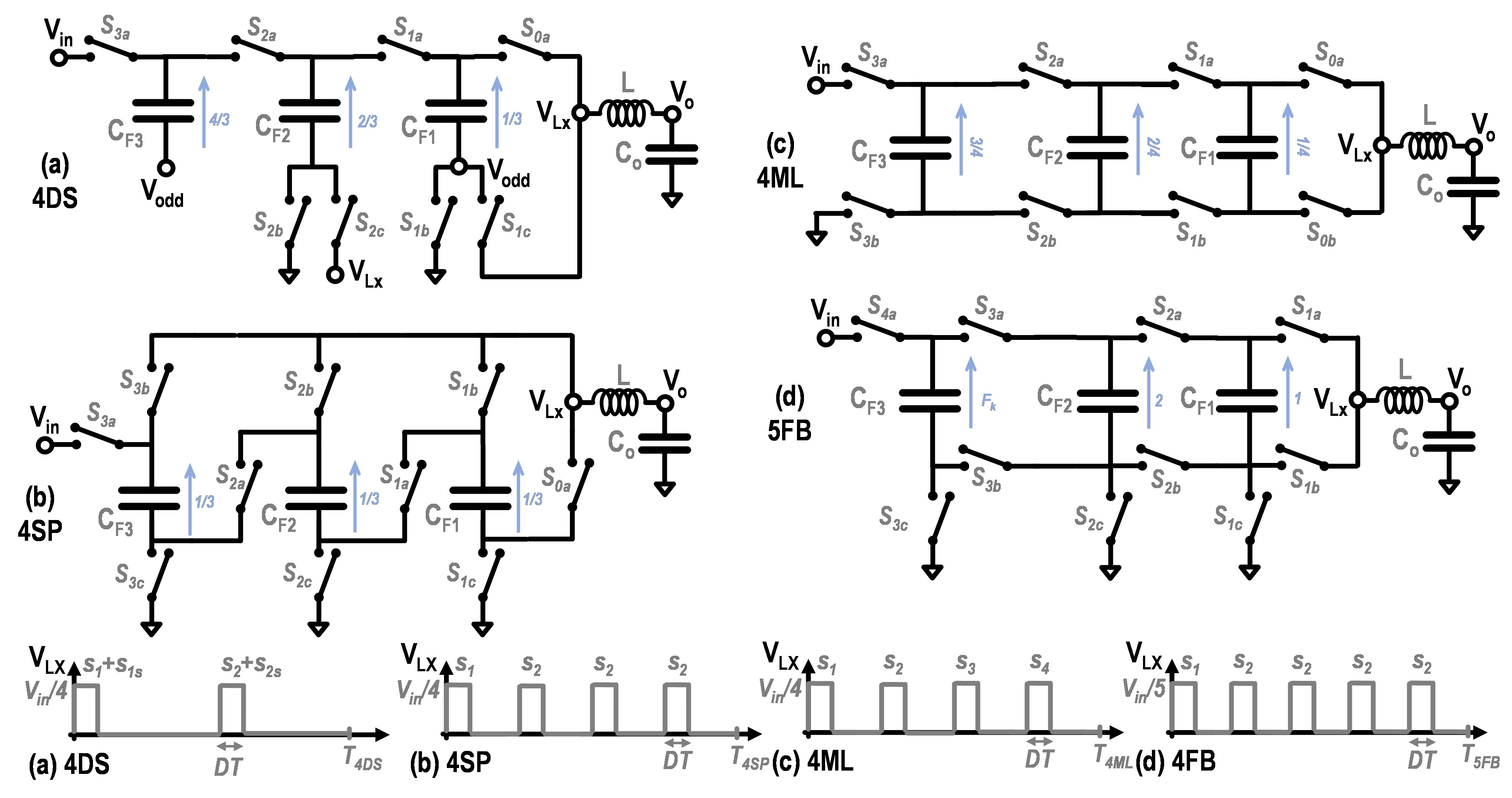}
    \caption{Selected Topologies: Schematic Diagrams for N=4}
    \label{fig:2}
    \end{center}
\end{figure*}

\section{PROPOSED COMPARISON FRAMEWORK}

\subsection{Assumptions and Notations}
Even though we attempted to find a generic method, our approach is only valid in continuous mode, in steady-state, without core saturation or hysteresis effects, with a single inductor $L$ at the output, and in fully soft-charging operation. We treat the HSCC operation in buck-mode far below the resonance \cite{yousefzadeh_three-level_2006, lei_split-phase_2016}. We focus on single-phase HSCC, but the methodology could be extended to $N$-phase. The approach also assumes that the inductor series resistance (DCR) does not vary significantly in the switching frequency range studied here (approximately ten times). For simplicity, all flying capacitors share the same value, $C_F$. All symbols used in the paper are defined and explained throughout the manuscript, and are also summarized in the Appendix.

\subsection{Method to size the converter at iso-loss and iso-ripple}

The method proposed in this study compares the performance of specific HSCC topologies with that of a 2-level single-phase buck converter (1B) at an equivalent power dissipation level, considering inductor current and output voltage ripples.

The losses considered in this analysis include the conduction loss of the switches ($P_{cond}$), gate driving loss ($P_{drive}$), and inductor conduction loss from both DCR and inductor ripple ($P_{ind}$). The total converter loss is expressed as:
\begin{equation}\label{E:1}
\begin{split}
\widetilde{P_{loss,k}} = \widetilde{P_{cond,k}} + \widetilde{P_{drive,k}}+\widetilde{P_{ind,k}}  
\\
=\widetilde{R_{o,k}} \widetilde{I_{o}^2} + \widetilde{E_{dr,k}} \widetilde{F_{k}} + R_L \widetilde{I_{o}^2}(1+\frac{\varepsilon}{12})
\end{split}
\end{equation}
where $k$ is the topology index, $R_o$ is the equivalent output impedance, $I_o$ is the output current, $F$ is the switching frequency, $E_{dr}$ is the total energy required to commute all switches during one cycle, $R_L$ is the DCR of the ouput inductor, and $\varepsilon$ is the relative inductor current ripple (relative to $I_{o}$). The purpose of using the tilde notation is to differentiate between the normalized values with respect to the baseline topology, 1B (which are represented without tilde) and the dimensional values (which are represented with tilde).

As we always compare topology $k$ to 1B for normalization purposes, the comparison procedure follows the steps below for a given VCR, denoted by $M=V_{o}/V_{i}$:

\begin{itemize}
\item Find the total switch area, $A$, to achieve the \textbf{same output impedance as 1B} (same $\widetilde{P_{cond}}$), using the well-known relationship between on-state resistance, area, and voltage blocking requirement (see II.D).
\item Adapt the switching frequency, $F$, to \textbf{equalize the switching loss to 1B} $\widetilde{P_{drive}}$, using the well-adopted relationship between on-state resistance, gate capacitance, and blocking voltage (see II.E).
\item Adapt the inductor and capacitor values to obtain the \textbf{same relative inductor current and output voltage ripples as 1B}, for equalizing $\widetilde{P_{ind}}$. The flying capacitors $C_{F}$ value is also derived accordingly to operate far below the resonance.
\end{itemize}

As outputs, we obtain the switch area $A_{k}$, the switching frequency $F$, and inductor value $L$, all relative to 1B. From these values, we can deduce the gain in passive size (inductor and capacitor) and bandwidth compared to the baseline 1B buck converter.

Compared to previous works, except for \cite{lei_analytical_2015}, it is important to notice that the couple \{$A$,$F$\} is adapted to achieve the same power efficiency, leading to adapting the passive triptych \{$L$,$C_{F}$,$C_{o}$\}. In \cite{lei_analytical_2015}, a similar comparison procedure is proposed, but we extend the case to asymmetric switch sizing, adapting the conductance of each switch to minimize the conduction loss at the targeted $M$.

The targeted VCR, $M$, is not necessarily the inherent VCR ($N$) of the SCC network placed before the inductor ($M \ne N$). Our analysis is done to compare HSCC with regulation capabilities. The voltage regulation is achieved by adding an extra state, denoted by $G$ here, that sets the left inductor terminal ($V_{Lx}$) to ground (Fig. \ref{fig:1}).

\subsection{Active-devices scaling models}

Before explaining in detail each step, let's discuss the trade-off between switch performance and voltage rating. Common cost-scaling models have already been introduced to create a relationship between the size, on-state resistance, switching energy, and voltage rating \cite{baliga_power_1989, stauth_pathways_2018, liu_97_2020}. Depending on the scenario (cascade arrangement, constant field, high-voltage), the specific resistance $R_{sp}$ (resistance-area product), and switching energy per unit area $E_{dr}$ have a relationship with the blocking voltage:
\begin{equation}\label{E:2}
\widetilde{R_{sp}} \propto \widetilde{V_{s}}^\alpha  ; \widetilde{E_{dr}} \propto \widetilde{V_{s}}^\beta
\end{equation}
Where $\alpha$ and $\beta$ vary from 0 to 2, and $V_s$ is the blocking voltage withstand by the switch.

\subsection{Step 1: switch area determination}

The first step consists of finding the minimal switch area, $\widetilde{A_{k,min}}$, for a given output impedance, which is fixed by the 1B reference ($\widetilde{R_{o,1B}}$):
\begin{equation}
\widetilde{A_{k,min}}=\arg\min_{R_{i}}\widetilde{A_k}
\end{equation}
where $R_{i}$ are the on-state resistance of the $i^{th}$ switch when $\widetilde{A_k}$ is minimal.

Inspired by \cite{seeman_analysis_2008}, $\widetilde{A_k}$ and $\widetilde{R_{o,k}}$ are linked by:
\begin{equation}\label{E:3}
\widetilde{A_k}= \sum_{i}^{switch}\frac{\widetilde{V_{s,i}}^\alpha }{\widetilde{R_i}} ; \widetilde{R_{o,k}} = \sum_{i}^{switch} R_{i} C_{i} (M) \widetilde{I_o}^2
\end{equation}
where $V_{s,i}$ is the blocking voltage of the $i$th switch. $C_{i}$ is the fraction of output RMS current flowing in the $i^{th}$ switch (explained later).

To be agnostic to the output current $\widetilde{I_o}$, input voltage $\widetilde{V_{in}}$, output impedance $\widetilde{R_o}$, and resistance-area product $\widetilde{R_{sp}}$, a normalization process is applied by dividing the output impedance of the studied topology $k$ by 1B:

\begin{equation} \label{E:4}
A_k= \frac{\widetilde{A_k(R_o)}}{\widetilde{A_{1B}} (R_o)} = f(C_i, V_{s,i}, \alpha)
\end{equation}
Here, $\widetilde{A_k}(R_{o})$ is the switch area when the output impedance is equal to $R_o$, and $A_k$ is the normalized area.

Compared to previous works, we choose to define a “current multiplier”, $C_{i}$, instead of a “charge multiplier”, $q_{i}$ \cite{seeman_analysis_2008}, as soft-charging operation forces the current flow not a charge quantity. To determine $C_i$, the current flow in each switch has to be examined:

\begin{equation} \label{E:5}
\widetilde{I_{sw,i}} = C_i(M) \widetilde{I_o}
\end{equation}
Here, $\widetilde{I_{sw,i}}$ is the RMS current flowing in the $i^{th}$ switch during one switching period (1/$F$).

 Moreover, the comparison is not performed at the same switch area, $A_{k}$, as usual, but at the same output impedance, $R_{o}$, in order to equalize conduction loss $P_{cond}$. Since the conduction loss is not frequency-dependent as in an SCC, the switching frequency, $F$, does not play a role here and is only determined in step 2. The minimal $A_{k}$ and optimal on-state resistance repartition $R_{i}$ can be found using numerical optimization, such as exhaustive search, or analytical resolution such as Lagrange multiplier.
 
\begin{table*}[]
\caption{Topological Parameters of the Selected HSCC}
\resizebox{\textwidth}{!}{

\begin{tabular}{|c|c|c|c|c|c|c|ccc|c|}
\hline
 & \#cap. & \#switch & m & d & p & s & \multicolumn{1}{c|}{$C^2$} & \multicolumn{1}{c|}{$S$} & $V_s$ & $V_c$ \\ \cline{2-11} 
Topo & \multicolumn{1}{l|}{} & \multicolumn{1}{l|}{} & $V_{Lx}$/$\widetilde{V_{in}}$ & $D/M$ & $V_{Lx}$   pulses & \multicolumn{1}{l|}{} & \multicolumn{1}{c|}{Current Multiplier} & \multicolumn{1}{c|}{Switch Activity} & Blocking Volt. & Cap. DC Volt. \\ \cline{3-3} \cline{8-11} 
 & \multicolumn{1}{l|}{} & \multicolumn{1}{l|}{} &  &  &  & \multicolumn{1}{l|}{} & \multicolumn{3}{c|}{S0a;S0b;S1a;S1b;S1c;S2a;S2b;S2c;S3a;S3b;S3c;S4a} & $C_{F1}$;$C_{F2}$;$C_{F3}$ \\ \hline
1B & 0 & 2 & 1 & 1 & 1 & - & \multicolumn{1}{c|}{M,1-M} & \multicolumn{1}{c|}{1;1} & 1;1 & - \\ \hline
2ML & 1 & 4 & 1/2 & 1 & 2 & 1 & \multicolumn{1}{c|}{M;1-M;M;1-M} & \multicolumn{1}{c|}{1;1;1;1} & 1/2;1/2;1/2;1/2 & 1/2 \\ \hline
3ML & 2 & 6 & 1/3 & 1 & 3 & 2 & \multicolumn{1}{c|}{M;1-M;M;1-M;M;1-M} & \multicolumn{1}{c|}{1;1;1;1;1;1} & 1/3;1/3;1/3;1/3;1/3;1/3 & 1/3;2/3 \\ \cline{1-1} \cline{3-11} 
3SP &  & 7 & 1/3 & 1 & 3 & 2 & \multicolumn{1}{c|}{1-2M;M;M/2;1-5/2M;M;M/2;M/2} & \multicolumn{1}{c|}{2;1;2;1;1;1;2} & 1/3;1/3;1/3;1/3;2/3;2/3;2/3 & 1/3;1/3 \\ \cline{1-1} \cline{3-11} 
3FB &  & 7 & 1/3 & 1 & 3 & 2 & \multicolumn{1}{c|}{1-5/2M;2M;1-2M;M;M;M/2;M/2} & \multicolumn{1}{c|}{1;2;2;1;1;1;2} & 1/3;1/3;1/3;1/3;2/3;2/3;1/3 & 1/3;2/3 \\ \cline{1-1} \cline{3-11} 
3DS &  & 7 & 1/3 & 2/3 & 2 & 2 & \multicolumn{1}{c|}{3/4M;M;1/4;1/4+M/4;3/4M;1/4+M/4;1/4} & \multicolumn{1}{c|}{2;1;1;2;1;1;1} & 1/3;2/3;1/3;1/3;1/3;1/3;1/3 & 1/3;2/3 \\ \hline
4ML & 3 & 8 & 1/4 & 1 & 4 & 2 & \multicolumn{1}{c|}{M;1-M;M;1-M;M;1-M;M;1-M} & \multicolumn{1}{c|}{1;1;1;1;1;1;1;1} & 1/4;1/4;1/4;1/4;1/4;1/4;1/4;1/4 & 1/4;1/2;3/4 \\ \cline{1-1} \cline{3-11} 
\multirow{2}{*}{4SP} &  & 10 & 1/4 & 1 & 4 & 3 & \multicolumn{1}{c|}{1-3M;M;M/3;1-11/3M;} & \multicolumn{1}{l|}{3;1;3;1;1;1;3;1;1;3} & 1/4;1/4;1/4;1/4;1/4; & 1/4;1/4;1/4 \\
 &  &  &  &  &  &  & \multicolumn{1}{c|}{;M/3;M/3;M;M/3;M/3} & \multicolumn{1}{l|}{} & 1/2;1/2;3/4;3/4;3/4 &  \\ \cline{1-1} \cline{3-11} 
\multirow{2}{*}{5FB} &  & 10 & 1/5 & 1 & 5 & 2/3 & \multicolumn{1}{c|}{1-14/3M;3M;1-3M;M/2;} & \multicolumn{1}{l|}{2;3;3;2;2;3;3;3;2;2} & 1/5;1/5;1/5;1/5;2/5; & 1/5;2/3;3/5 \\
 &  &  &  &  &  &  & \multicolumn{1}{c|}{M;M/3;M/3;M/3;M/2;M/2} & \multicolumn{1}{l|}{} & 2/5;2/5;3/5;3/5;2/5 &  \\ \cline{1-1} \cline{3-11} 
\multirow{2}{*}{4DS} &  & 8 & 1/4 & 2 & 2 & 2 & \multicolumn{1}{c|}{2/3M;2/3M;1/4+M;1/4+M; 2/3M} & \multicolumn{1}{l|}{1;1;1;1;1;1;1;1} & 1/4;1/2;1/4;1/4;1/2; & 1/4;1/2;3/4 \\
 &  &  &  &  &  &  & \multicolumn{1}{c|}{1/4-M/3;1/4-M/3;2/3M} & \multicolumn{1}{l|}{} & 1/4;1/4;1/4 &  \\ \hline
\end{tabular}
}
\label{tab:topological_parameters}
\end{table*}

\subsection{Step 2: Switching Frequency Determination}

The normalized energy required to switch all the switches, $E_{dr}$, is determined in the previous step using:

\begin{equation}
E_{dr,k}= \frac{\widetilde{E_{dr,k}}}{\widetilde{E_{dr,1B}}} = \sum_{i}^{\#switch} S_{i}\frac{V_{s,i}^{(\alpha+\beta)}}{R_{i}}
\end{equation}
Here, $S_{i}$ is the relative switching rate (referenced to $F$).

As $F$ is the relative switching frequency considering $F_{1B}$, which is equal to unity, it can be expressed as:

\begin{equation}\label{E:7}
F_k=\frac{\widetilde{F_k}}{\widetilde{F_{1B}}}=\frac{1}{E_{dr,k}}
\end{equation}

This frequency keeps the same driving loss as that in 1B. Similar to other normalized parameters, $F_k$ does not depend on $\widetilde{V_{in}}$ or output power as the normalization is done compared to 1B, which delivers the same power under the same $\widetilde{V_{in}}$.

\subsection{Step 3: Passive values determination}

When SCC networks impose the voltage at the switching node, $V_{Lx}$, the inductor current ripple can be obtained using periodic steady-state constraint as:

\begin{equation}
\widetilde{\Delta I_{L,k}}= \frac{\widetilde{V_L} \widetilde{T_s}}{\widetilde{L_k}}=\widetilde{D_k}\frac{\widetilde{V_{Lx,k}} - \widetilde{V_o}}{\widetilde{L_k}\widetilde{F_k}}
\end{equation}
Here, $\widetilde{V_{lx,k}}$ is the voltage generated on the output of the switched-capacitor structure (Fig. 1), $\widetilde{D_k}$ is the duty cycle, and $T_s$ is the duration of one state.

To keep the method as generalized as possible, we perform a normalization and compare the ripple to 1B, meaning the current ripple in the inductor are the same than 1B,  as:

\begin{equation}
\frac{\widetilde{\Delta I_{L,k}}}{\widetilde{\Delta I_{L,1B}}}=\frac{d_k (m_k-M)}{L_k F_k (1-M)} =1
\end{equation}
Here, $d_k$ is the normalized duty cycle ratio $(\widetilde{D_k}/M)$, $m_k$ is defined by max($V_{Lx,k}$)/$\widetilde{V_{in}}$, and $d_{1B}=m_{1B}=1$ by previous definitions.

From topology inspection and the previous step determining $F_k$, the relative inductor value $L_{k}$ can be found. The output capacitor value, $C_{o,k}$, can be extracted in a similar manner and is given by:

\begin{equation}
C_{o,k}=\frac{1}{L_kF_k^2} \frac{d_k(m_k-M)}{p_k(1-M)}
\end{equation}
Here, $p_K$ is the number of pulses experienced by $V_{Lx}$ during one HSCC period.

Contrary to \cite{lei_analytical_2015} where voltage ripples across each $C_F$ is kept constant, the flying capacitor values here are deduced by keeping the operation out-of-resonance under the same SCC ratio by the following:

\begin{equation} \label{E:11}
C_{F,k}=\frac{s_{k} d_{k}^2}{L_{k} F_{k}^2}
\end{equation}
Here, $s_{k}$ is the relative equivalent flying capacitor forming by the SCC network in all states referred to $C_{F}$.

\subsection{Inductor volume and bandwidth}
\label{sect:ULandBW}

In this paper, we assume the inductor dictates the total volume, as is often the case in most practical designs \cite{lei_analytical_2015}, though future work will look at including the capacitor volume as well. The inductor volume can be evaluated by the stored energy in the inductor ($\epsilon$$<<$$1$, $\Delta V_{ci}$$<<$$V_{ci}$):
\begin{equation}
U_{L,k}=\frac{\widetilde{U_{L,k}}}{\widetilde{U_{L,1B}}}=L_{k}
\end{equation}

From passive values and similarly to \cite{yousefzadeh_three-level_2005}, the output filter corner frequency or the open-loop bandwidth can be deduced:
\begin{equation}
BW_k\propto \frac{1}{\sqrt{L_k C_{o,k}}}
\end{equation}

As in some applications, both the transient response and the inductor volume are two important design objectives, we also introduce an arbitrary FoM which multiplies $BW_{k}$ and $1/U_{L,k}$ to represent this trade-off.

\section{Topology Benchmark}

\subsection{Selected topologies}

In this paper, we have limited our analysis up to three flying capacitors to keep a reasonable number of passive devices, although the analysis could be easily extended to more. We have also only considered HSCCs presenting fully soft-charging operation with a single inductor at the output. This limits the selection of all common SCC topologies as they are not always compatible with soft-charging. Some proposed topologies presenting both soft-charging and hard-charging operations, e.g., for allowing dual-path converters \cite{jung_dual-path_2021}, are outside our scope. Configurations satisfying this requirement (without infinite flying capacitor values) have already been formally revealed in previous work \cite{lei_analytical_2015}: series-parallel (SP) and Fibonacci (FB). Dickson is not a natural soft-charging configuration, but it can be modified to be compatible by introducing split phases \cite{lei_split-phase_2016}, called here DS. Ladder and doubler topologies are not included in our benchmark as they are generally not amenable to soft-charging. According to our modeling assumptions, we compare SP, FB, and DS in the following. Additionally, traditional flying capacitor multilevel converters (ML) have been added to the benchmark as they also satisfy the constraint \cite{brooks_operation_2022,kim_fully-integrated_2012,lei_split-phase_2016}. These topologies align with the direct-conversion distinction proposed in \cite{kesarwani_resonant_2015}.

In this paper, the $N$ prefix is used for naming the $N$:1 natural SCC voltage conversion ratio. The number of flying capacitors is $N$-1 (except for 5FB, which follows the Fibonacci series). The $G$ state is the ground phase, meaning the phase where the inductor is demagnetized. The duration of the $G$ state is modulated to achieve the targeted voltage ratio $M$ from the natural $N$:1 ratio given by the flying capacitors network.

\subsection{Inputs determination using 4ML as an example}

In this section, we illustrate the determination of the topological parameters ($C$, $S$, $V_s$, $V_c$, $m$, $d$, $p$, $s$) for the 4-to-1 HSCC in multi-level configuration (4ML), which serves as an example. The schematic and four-state sequence are shown in Fig. 2 (c) and Table II, respectively.

In steady-state operation with ideal charge balancing (where state durations are assumed to be equal), the capacitor network provides a $V_{LX}$ pulse train toggling between $0$ and $\widetilde{v_{in}}/4$ ($m_k$=1/4). The relationship between VCR and duty cycle is $D=M$ ($d_k$=1) since there are four pulses with $\widetilde{V_{in}}/4$ amplitude and duration $D$ on $V_{Lx}$. Table II gives the normalized current flowing in each state (with respect to the output current). From this table, the RMS current flowing through each switch during one period is determined to obtain the $C$ vector (summarized in Table \ref{tab:topological_parameters}):

\begin{equation}
C_i=\sqrt{\frac{1}{T}\sum_j^{state}c_{i,j}^2T_j}=f(M)
\end{equation}
where $T_j$ is the duration of the $j^{th}$ phase, $c_{i,j}$ is the normalized current passing through the $i^{th}$ switch, $T$ is the period, and $f(M)$ is only a function of the VCR.

The $V_s$ vector, which illustrates the maximal blocking voltage experienced by each switch, is also determined from circuit inspection (Table I). In 4ML, the switching activity of all switches is equal unity as every switch commutes once (ON and OFF) during one period. Thus, $S$ is equal to the unity vector. As the topology has three flying capacitors, $V_c$ is a three-element vector where the $i^{th}$ value represents the DC voltage across the $i^{th}$ capacitor.

The parameters of all screened topologies are given in Table I. To reference switch labeling, three capacitor-based schematics are given in Fig. 2. The others follow the same logical naming but with lower number of switches. For charge balancing and soft-charging operation, the operation of the ten topologies presents some slight modifications compared to their baseline SCC counterparts. Again, Dickson NDS includes the splitting phase (called 1s, 2s) as introduced in \cite{lei_split-phase_2016}. The series-parallel NSP topology has one charging state (first phase) and $N$ discharging states (second state) for charge balancing purposes \cite{liu_97_2020}.

\begin{table*}[]
\caption{Normalized current ($I_{sw}/I_o$) flowing in each switch for each states in 10 selected topologies}
\resizebox{\textwidth}{!}{
\begin{tabular}{|c|cc|ccc|cccc|ccccc|ccc|ccc|ccc|ccc|ccccc|ccccc|}
\hline
topo & \multicolumn{2}{c|}{1B} & \multicolumn{3}{c|}{2ML} & \multicolumn{4}{c|}{3ML} & \multicolumn{5}{c|}{4ML} & \multicolumn{3}{c|}{3SP} & \multicolumn{3}{c|}{4SP} & \multicolumn{3}{c|}{3FB} & \multicolumn{3}{c|}{5FB} & \multicolumn{5}{c|}{3DS} & \multicolumn{5}{c|}{4DS} \\ \hline
state & \multicolumn{1}{c|}{1} & G & \multicolumn{1}{c|}{1} & \multicolumn{1}{c|}{2} & G & \multicolumn{1}{c|}{1} & \multicolumn{1}{c|}{2} & \multicolumn{1}{c|}{3} & G & \multicolumn{1}{c|}{1} & \multicolumn{1}{c|}{2} & \multicolumn{1}{c|}{3} & \multicolumn{1}{c|}{4} & G & \multicolumn{1}{c|}{1} & \multicolumn{1}{c|}{2} & G & \multicolumn{1}{c|}{1} & \multicolumn{1}{c|}{2} & G & \multicolumn{1}{c|}{1} & \multicolumn{1}{c|}{2} & G & \multicolumn{1}{c|}{1} & \multicolumn{1}{c|}{2} & G & \multicolumn{1}{c|}{1} & \multicolumn{1}{c|}{1s} & \multicolumn{1}{c|}{2} & \multicolumn{1}{c|}{2s} & G & \multicolumn{1}{c|}{1} & \multicolumn{1}{c|}{1s} & \multicolumn{1}{c|}{2} & \multicolumn{1}{c|}{2s} & G \\ \hline
Ts & \multicolumn{1}{c|}{D} & 1-D & \multicolumn{1}{c|}{D} & \multicolumn{1}{c|}{D} & 1-2D & \multicolumn{1}{c|}{D} & \multicolumn{1}{c|}{D} & \multicolumn{1}{c|}{D} & 1-3D & \multicolumn{1}{c|}{D} & \multicolumn{1}{c|}{D} & \multicolumn{1}{c|}{D} & \multicolumn{1}{c|}{D} & 1-4D & \multicolumn{1}{c|}{D} & \multicolumn{1}{c|}{2D} & 1-3D & \multicolumn{1}{c|}{D} & \multicolumn{1}{c|}{3D} & 1-4D & \multicolumn{1}{c|}{D} & \multicolumn{1}{c|}{2D} & 1-3D & \multicolumn{1}{c|}{D} & \multicolumn{1}{c|}{4D} & 1-5D & \multicolumn{1}{c|}{2/3D} & \multicolumn{1}{c|}{1/3D} & \multicolumn{1}{c|}{2/3D} & \multicolumn{1}{c|}{1/3D} & 1-2D & \multicolumn{1}{c|}{3/4D} & \multicolumn{1}{c|}{1/4D} & \multicolumn{1}{c|}{3/4D} & \multicolumn{1}{c|}{¼D} & 1-2D \\ \hline
0a & \multicolumn{1}{c|}{1} &  & \multicolumn{1}{c|}{1} & \multicolumn{1}{c|}{} &  & \multicolumn{1}{c|}{1} & \multicolumn{1}{c|}{} & \multicolumn{1}{c|}{} &  & \multicolumn{1}{c|}{1} & \multicolumn{1}{c|}{} & \multicolumn{1}{c|}{} & \multicolumn{1}{c|}{} &  & \multicolumn{1}{c|}{1} & \multicolumn{1}{c|}{} & 1 & \multicolumn{1}{c|}{1} & \multicolumn{1}{c|}{} & 1 & \multicolumn{1}{c|}{} & \multicolumn{1}{c|}{} &  & \multicolumn{1}{c|}{} & \multicolumn{1}{c|}{} &  & \multicolumn{1}{c|}{1/2} & \multicolumn{1}{c|}{} & \multicolumn{1}{c|}{} & \multicolumn{1}{c|}{1} &  & \multicolumn{1}{c|}{2/3} & \multicolumn{1}{c|}{} & \multicolumn{1}{c|}{} & \multicolumn{1}{c|}{} &  \\ \hline
0b & \multicolumn{1}{c|}{} & 1 & \multicolumn{1}{c|}{} & \multicolumn{1}{c|}{1} & 1 & \multicolumn{1}{c|}{} & \multicolumn{1}{c|}{1} & \multicolumn{1}{c|}{1} & 1 & \multicolumn{1}{c|}{} & \multicolumn{1}{c|}{1} & \multicolumn{1}{c|}{1} & \multicolumn{1}{c|}{1} & 1 & \multicolumn{1}{c|}{} & \multicolumn{1}{c|}{} &  & \multicolumn{1}{c|}{} & \multicolumn{1}{c|}{} &  & \multicolumn{1}{c|}{} & \multicolumn{1}{c|}{} &  & \multicolumn{1}{c|}{} & \multicolumn{1}{c|}{} &  & \multicolumn{1}{c|}{} & \multicolumn{1}{c|}{} & \multicolumn{1}{c|}{} & \multicolumn{1}{c|}{} &  & \multicolumn{1}{c|}{} & \multicolumn{1}{c|}{} & \multicolumn{1}{c|}{} & \multicolumn{1}{c|}{} &  \\ \hline
1a & \multicolumn{1}{c|}{} &  & \multicolumn{1}{c|}{} & \multicolumn{1}{c|}{1} &  & \multicolumn{1}{c|}{} & \multicolumn{1}{c|}{} & \multicolumn{1}{c|}{1} &  & \multicolumn{1}{c|}{} & \multicolumn{1}{c|}{} & \multicolumn{1}{c|}{1} & \multicolumn{1}{c|}{} &  & \multicolumn{1}{c|}{1} & \multicolumn{1}{c|}{} &  & \multicolumn{1}{c|}{1} & \multicolumn{1}{c|}{} &  & \multicolumn{1}{c|}{} & \multicolumn{1}{c|}{1/2} & 1 & \multicolumn{1}{c|}{} & \multicolumn{1}{c|}{1/3} & 1 & \multicolumn{1}{c|}{} & \multicolumn{1}{c|}{} & \multicolumn{1}{c|}{1} & \multicolumn{1}{c|}{} &  & \multicolumn{1}{c|}{} & \multicolumn{1}{c|}{} & \multicolumn{1}{c|}{1/3} & \multicolumn{1}{c|}{1} &  \\ \hline
1b & \multicolumn{1}{c|}{} &  & \multicolumn{1}{c|}{1} & \multicolumn{1}{c|}{} & 1 & \multicolumn{1}{c|}{1} & \multicolumn{1}{c|}{1} & \multicolumn{1}{c|}{} & 1 & \multicolumn{1}{c|}{1} & \multicolumn{1}{c|}{1} & \multicolumn{1}{c|}{} & \multicolumn{1}{c|}{1} & 1 & \multicolumn{1}{c|}{} & \multicolumn{1}{c|}{1/2} &  & \multicolumn{1}{c|}{} & \multicolumn{1}{c|}{1/3} &  & \multicolumn{1}{c|}{} & \multicolumn{1}{c|}{1} &  & \multicolumn{1}{c|}{} & \multicolumn{1}{c|}{1} &  & \multicolumn{1}{c|}{1/2} & \multicolumn{1}{c|}{} & \multicolumn{1}{c|}{} & \multicolumn{1}{c|}{1} & 1/2 & \multicolumn{1}{c|}{1} & \multicolumn{1}{c|}{1} & \multicolumn{1}{c|}{} & \multicolumn{1}{c|}{} & 1/2 \\ \hline
1c & \multicolumn{1}{c|}{} &  & \multicolumn{1}{c|}{} & \multicolumn{1}{c|}{} &  & \multicolumn{1}{c|}{} & \multicolumn{1}{c|}{} & \multicolumn{1}{c|}{} &  & \multicolumn{1}{c|}{} & \multicolumn{1}{c|}{} & \multicolumn{1}{c|}{} & \multicolumn{1}{c|}{} &  & \multicolumn{1}{c|}{} & \multicolumn{1}{c|}{1/2} & 1 & \multicolumn{1}{c|}{} & \multicolumn{1}{c|}{1/3} & 1 & \multicolumn{1}{c|}{1} & \multicolumn{1}{c|}{} & 1 & \multicolumn{1}{c|}{1} & \multicolumn{1}{c|}{} & 1 & \multicolumn{1}{c|}{} & \multicolumn{1}{c|}{} & \multicolumn{1}{c|}{1} & \multicolumn{1}{c|}{} & 1/2 & \multicolumn{1}{c|}{} & \multicolumn{1}{c|}{} & \multicolumn{1}{c|}{1} & \multicolumn{1}{c|}{1} & 1/2 \\ \hline
2a & \multicolumn{1}{c|}{} &  & \multicolumn{1}{c|}{} & \multicolumn{1}{c|}{} &  & \multicolumn{1}{c|}{} & \multicolumn{1}{c|}{1} & \multicolumn{1}{c|}{} &  & \multicolumn{1}{c|}{} & \multicolumn{1}{c|}{} & \multicolumn{1}{c|}{} & \multicolumn{1}{c|}{1} &  & \multicolumn{1}{c|}{1} & \multicolumn{1}{c|}{} &  & \multicolumn{1}{c|}{1} & \multicolumn{1}{c|}{} &  & \multicolumn{1}{c|}{1} & \multicolumn{1}{c|}{} &  & \multicolumn{1}{c|}{1/2} & \multicolumn{1}{c|}{} &  & \multicolumn{1}{c|}{1/2} & \multicolumn{1}{c|}{1} & \multicolumn{1}{c|}{} & \multicolumn{1}{c|}{} &  & \multicolumn{1}{c|}{1/3} & \multicolumn{1}{c|}{1} & \multicolumn{1}{c|}{} & \multicolumn{1}{c|}{} &  \\ \hline
2b & \multicolumn{1}{c|}{} &  & \multicolumn{1}{c|}{} & \multicolumn{1}{c|}{} &  & \multicolumn{1}{c|}{1} & \multicolumn{1}{c|}{} & \multicolumn{1}{c|}{1} & 1 & \multicolumn{1}{c|}{1} & \multicolumn{1}{c|}{1} & \multicolumn{1}{c|}{1} & \multicolumn{1}{c|}{} & 1 & \multicolumn{1}{c|}{} & \multicolumn{1}{c|}{1/2} &  & \multicolumn{1}{c|}{} & \multicolumn{1}{c|}{1/3} &  & \multicolumn{1}{c|}{1} & \multicolumn{1}{c|}{} &  & \multicolumn{1}{c|}{1} & \multicolumn{1}{c|}{} &  & \multicolumn{1}{c|}{} & \multicolumn{1}{c|}{} & \multicolumn{1}{c|}{1} & \multicolumn{1}{c|}{} & 1/2 & \multicolumn{1}{c|}{} & \multicolumn{1}{c|}{} & \multicolumn{1}{c|}{1/3} & \multicolumn{1}{c|}{1} & 1/2 \\ \hline
2c & \multicolumn{1}{c|}{} &  & \multicolumn{1}{c|}{} & \multicolumn{1}{c|}{} &  & \multicolumn{1}{c|}{} & \multicolumn{1}{c|}{} & \multicolumn{1}{c|}{} &  & \multicolumn{1}{c|}{} & \multicolumn{1}{c|}{} & \multicolumn{1}{c|}{} & \multicolumn{1}{c|}{} &  & \multicolumn{1}{c|}{} & \multicolumn{1}{c|}{1/2} &  & \multicolumn{1}{c|}{} & \multicolumn{1}{c|}{1/3} &  & \multicolumn{1}{c|}{} & \multicolumn{1}{c|}{1/2} &  & \multicolumn{1}{c|}{} & \multicolumn{1}{c|}{1/3} &  & \multicolumn{1}{c|}{1/2} & \multicolumn{1}{c|}{1} & \multicolumn{1}{c|}{} & \multicolumn{1}{c|}{} & 1/2 & \multicolumn{1}{c|}{1/3} & \multicolumn{1}{c|}{1} & \multicolumn{1}{c|}{} & \multicolumn{1}{c|}{} & 1/2 \\ \hline
3a & \multicolumn{1}{c|}{} &  & \multicolumn{1}{c|}{} & \multicolumn{1}{c|}{} &  & \multicolumn{1}{c|}{} & \multicolumn{1}{c|}{} & \multicolumn{1}{c|}{} &  & \multicolumn{1}{c|}{} & \multicolumn{1}{c|}{1} & \multicolumn{1}{c|}{} & \multicolumn{1}{c|}{} &  & \multicolumn{1}{c|}{} & \multicolumn{1}{c|}{} &  & \multicolumn{1}{c|}{1} & \multicolumn{1}{c|}{} &  & \multicolumn{1}{c|}{} & \multicolumn{1}{c|}{1/2} &  & \multicolumn{1}{c|}{} & \multicolumn{1}{c|}{1/3} &  & \multicolumn{1}{c|}{} & \multicolumn{1}{c|}{} & \multicolumn{1}{c|}{} & \multicolumn{1}{c|}{} &  & \multicolumn{1}{c|}{} & \multicolumn{1}{c|}{} & \multicolumn{1}{c|}{2/3} & \multicolumn{1}{c|}{} &  \\ \hline
3b & \multicolumn{1}{c|}{} &  & \multicolumn{1}{c|}{} & \multicolumn{1}{c|}{} &  & \multicolumn{1}{c|}{} & \multicolumn{1}{c|}{} & \multicolumn{1}{c|}{} &  & \multicolumn{1}{c|}{1} & \multicolumn{1}{c|}{} & \multicolumn{1}{c|}{1} & \multicolumn{1}{c|}{1} & 1 & \multicolumn{1}{c|}{} & \multicolumn{1}{c|}{} &  & \multicolumn{1}{c|}{} & \multicolumn{1}{c|}{1/3} &  & \multicolumn{1}{c|}{} & \multicolumn{1}{c|}{} &  & \multicolumn{1}{c|}{} & \multicolumn{1}{c|}{1/3} &  & \multicolumn{1}{c|}{} & \multicolumn{1}{c|}{} & \multicolumn{1}{c|}{} & \multicolumn{1}{c|}{} &  & \multicolumn{1}{c|}{} & \multicolumn{1}{c|}{} & \multicolumn{1}{c|}{} & \multicolumn{1}{c|}{} &  \\ \hline
3c & \multicolumn{1}{c|}{} &  & \multicolumn{1}{c|}{} & \multicolumn{1}{c|}{} &  & \multicolumn{1}{c|}{} & \multicolumn{1}{c|}{} & \multicolumn{1}{c|}{} &  & \multicolumn{1}{c|}{} & \multicolumn{1}{c|}{} & \multicolumn{1}{c|}{} & \multicolumn{1}{c|}{} &  & \multicolumn{1}{c|}{} & \multicolumn{1}{c|}{} &  & \multicolumn{1}{c|}{} & \multicolumn{1}{c|}{1/3} &  & \multicolumn{1}{c|}{} & \multicolumn{1}{c|}{} &  & \multicolumn{1}{c|}{1/2} & \multicolumn{1}{c|}{} &  & \multicolumn{1}{c|}{} & \multicolumn{1}{c|}{} & \multicolumn{1}{c|}{} & \multicolumn{1}{c|}{} &  & \multicolumn{1}{c|}{} & \multicolumn{1}{c|}{} & \multicolumn{1}{c|}{} & \multicolumn{1}{c|}{} &  \\ \hline
4a & \multicolumn{1}{c|}{} &  & \multicolumn{1}{c|}{} & \multicolumn{1}{c|}{} &  & \multicolumn{1}{c|}{} & \multicolumn{1}{c|}{} & \multicolumn{1}{c|}{} &  & \multicolumn{1}{c|}{} & \multicolumn{1}{c|}{} & \multicolumn{1}{c|}{} & \multicolumn{1}{c|}{} &  & \multicolumn{1}{c|}{} & \multicolumn{1}{c|}{} &  & \multicolumn{1}{c|}{} & \multicolumn{1}{c|}{} &  & \multicolumn{1}{c|}{} & \multicolumn{1}{c|}{} &  & \multicolumn{1}{c|}{1/2} & \multicolumn{1}{c|}{} &  & \multicolumn{1}{c|}{} & \multicolumn{1}{c|}{} & \multicolumn{1}{c|}{} & \multicolumn{1}{c|}{} &  & \multicolumn{1}{c|}{} & \multicolumn{1}{c|}{} & \multicolumn{1}{c|}{} & \multicolumn{1}{c|}{} &  \\ \hline
\end{tabular}
}
\label{tab:states_description}
\end{table*}

\subsection{Design parameters determination}

\begin{figure}
    \begin{center}
    \captionsetup{justification=centering, belowskip=-20pt}
    \includegraphics[width=3.5in]{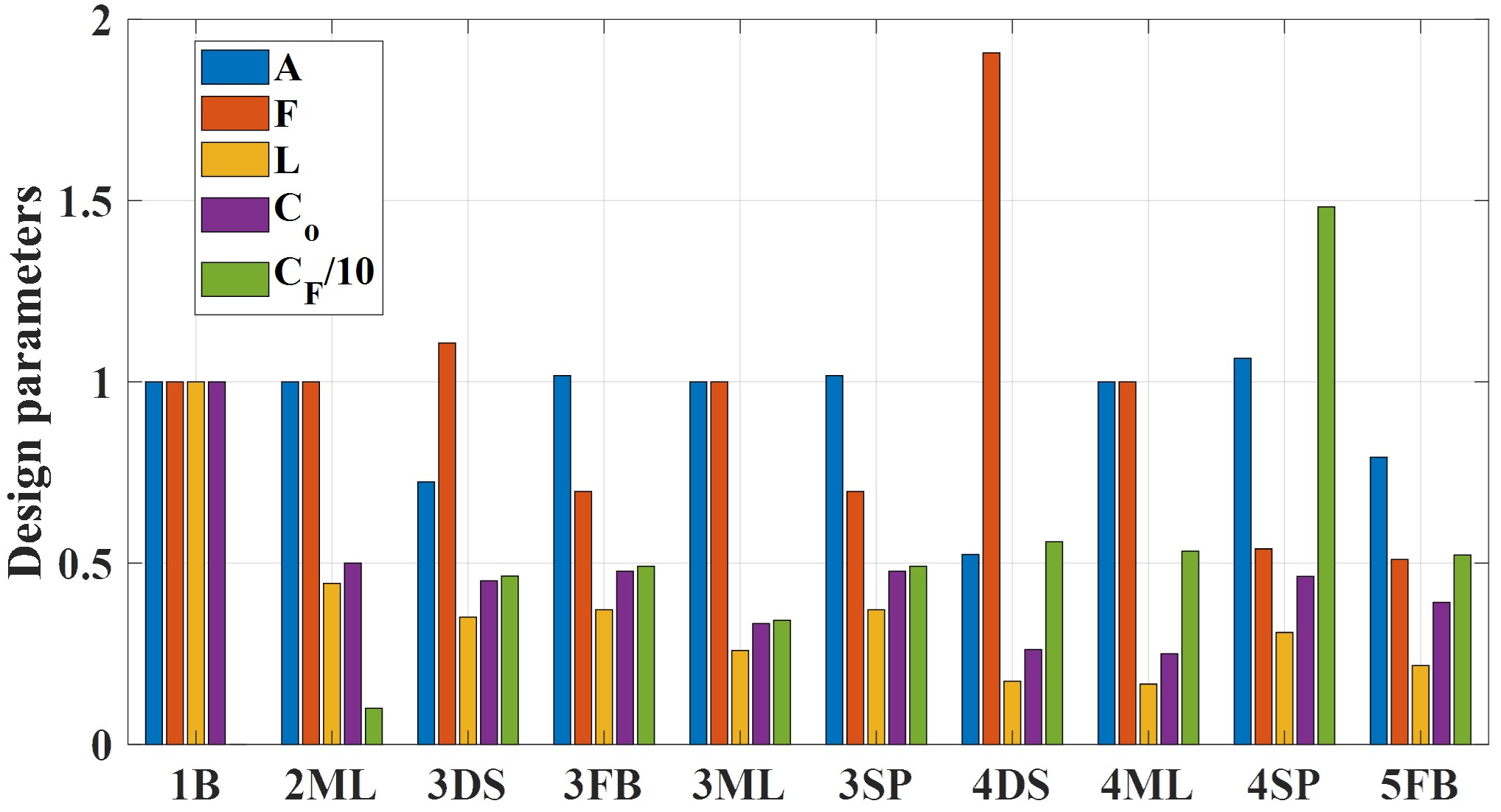}
    \caption{Optimal Design Parameters for a given set of input variables (M=0.1,$\alpha$=2,$\beta$=1)}
    \label{fig:3}
    \end{center}
\end{figure}

To obtain the same power efficiency and voltage output ripple as 1B for a given VCR ($M$), the design parameters of the ten aforementioned topologies, i.e., switch area $A$, switching frequency $F$, and passive component values ($L$, $C_{o}$, $C_{F}$), have been determined numerically following the three steps described in Section II. Fig. \ref{fig:3} depicts these variables for a particular set of input variables, consistent with the $G-V^2$ method developed in \cite{seeman_analysis_2008}. It should be noted that the conclusions drawn here depend on these input parameters and are specific to the particular context introduced previously (see the discussion section for more insight).

In step 1, the minimization of switch area $A$ needs to be performed for the targeted VCR $M$ (where $M\neq1/N$) and the partitioning of switches among $R_i$ stages. For instance, in DS topology, switches $S_{i,b}$ and $S_{i,c}$ are more bulky than $S_{i,a}$ (where $D<<1$). The general trend is that as $N$ increases (more flying capacitors), the values of inductor $L$ and output capacitor $C_o$ decrease, while the flying capacitor $C_F$ value tends to increase.

For a given switching activity $S$, the switching frequency $F$ (shown as orange bars in Fig. \ref{fig:3}) is inversely proportional to $A$, as given by Equation (\ref{E:7}). However, the FB and SP topologies have a significant drawback as the extra phases introduce multiple commutations in one cycle ($S_i=2$ or $3$ in some switches, per Table I). To maintain the same relative inductor current ripple (shown as yellow bars in Fig. \ref{fig:3}), Dickson topologies require longer pulse widths (higher $d$), which is slightly compensated by higher $F$ than others. The conclusion drawn in \cite{lei_analytical_2015} is that ML and DS have similar inductor values, which is similar to our findings. However, in contrast to \cite{lei_analytical_2015}, we observed that the value of flying capacitors in ML and DS is similar despite the differences in their sizing procedures.

Regarding the output capacitor $C_o$, topologies with more pulses ($p$) have an advantage for a given $N$. However, fewer $V_{Lx}$ pulses (low $p$) negatively impact the DS topology, but higher $F$ and lower $L$ values compensate to achieve a similar $C_o$ value than others. Lastly, the value of flying capacitor $C_F$ depends on the inductor $L$, switching frequency $F$, and the $s$ and $d$ parameters, as shown in (\ref{E:11}), to operate below the self-resonant frequency by the LC resonator formed by $L$ and the equivalent flying capacitors in the network.

In conclusion, for a given number of flying capacitors, the ML topology achieves the best reduction gain of the inductor value while maintaining the lowest capacitor values. It should be noted that the conclusions drawn here depend on the specific parameters and the context introduced previously, and should not be taken as a general answer for discarding certain topologies. Instead, the objective here is to provide a numerical example to showcase the power of the framework to benchmark different HSCCs.

\subsection{Inductor volume and bandwidth comparison}

\begin{figure}
    \begin{center}
    \captionsetup{justification=centering, belowskip=-20pt}
    \includegraphics[width=3.5in]{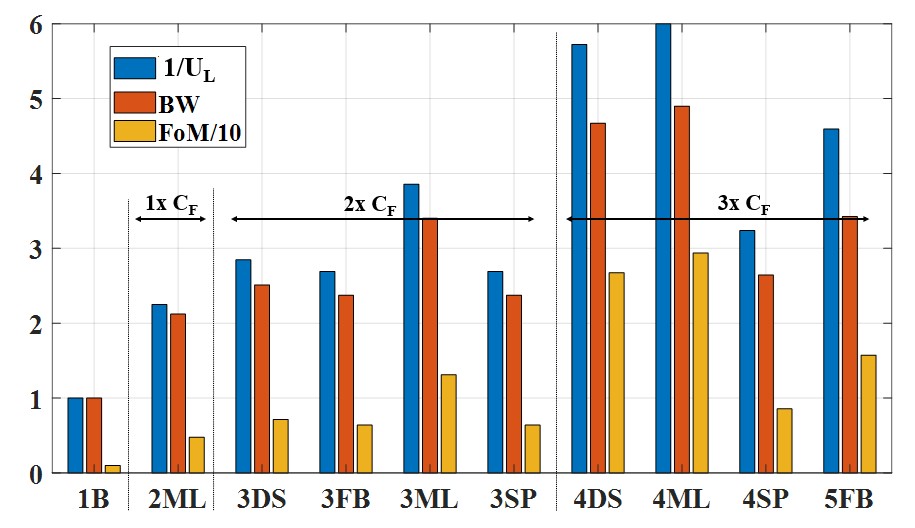}
    \caption{Passive volume, open-loop bandwidth and FoM comparison  for a given set of input variables (M=0.1,$\alpha$=2,$\beta$=1).}
    \label{fig:4}
    \end{center}
\end{figure}

As described in \ref{sect:ULandBW}, the inductor volume is deduced from the design variable values. In Fig.~\ref{fig:4}, inductor volume gain ($1/U_L$) tends to decrease with $N$. The 4DS and 4ML topologies are the best ones, reducing the volume by around 6x compared to 1B.

Figure~\ref{fig:4} also shows the comparison of the open-loop bandwidth (orange bar) and previously introduced FoM in II.G. There is a clear benefit for increasing $N$ to achieve a better bandwidth as almost all topologies tend to decrease the output filter values \{$L, C_{o}$\}. The FoM (yellows bars) shows the double benefits in volume and open-loop bandwidth in increasing the number of flying capacitors.

\section{Discussion}

In this paper, we present a comparison framework that considers multiple constraints, including the same power loss and same ripples, for evaluating the relative performance of different soft-charging single-inductor hybrid switched-capacitor (HSCC) topologies using only topology-dependent parameters and a few application-dependent input variables. It is important to note that the conclusions drawn in this study are only valid for non-resonant HSCCs, as resonant HSCCs have specific loss mechanisms \cite{ye_modeling_2022}. Some HSCC such as multiple inductors \cite{das_analysis_2021} placed at the input \cite{seo_s-hybrid_2020} or in the middle \cite{hardy_115_2023} of capacitor network are not covered in this paper as the loss mechanism has to be reconsidered and will be studied in future work.

Our framework establishes a clear relationship between the intrinsic switch stress (represented by the $A$ value) and the volume of the inductor, which is determined by the switching frequency $(F\propto 1/A)$ and topological relative parameters $(d, m, p)$. The benefit of this framework is exemplified by the results obtained for the DS family of topologies. The Dickson topology is widely recognized for its superior active utilization, leading to the lowest output impedance in the fast switching limit (FSL) \cite{seeman_analysis_2008}. This paper confirms the advantage of the Dickson topology in terms of minimizing the switch area, as depicted by the blue bars in Fig. \ref{fig:3}, which show a significant reduction compared to other topologies, reaching up to 50\%. However, we also demonstrate that when additional constraints, as mentioned earlier, are considered, and when splitting phases are incorporated into the sizing process, the Dickson topology's superiority is diminished, particularly when compared to multi-level topologies (ML).

\begin{figure}
    \begin{center}
    \captionsetup{justification=centering, belowskip=-20pt}
    \includegraphics[width=3.5in]{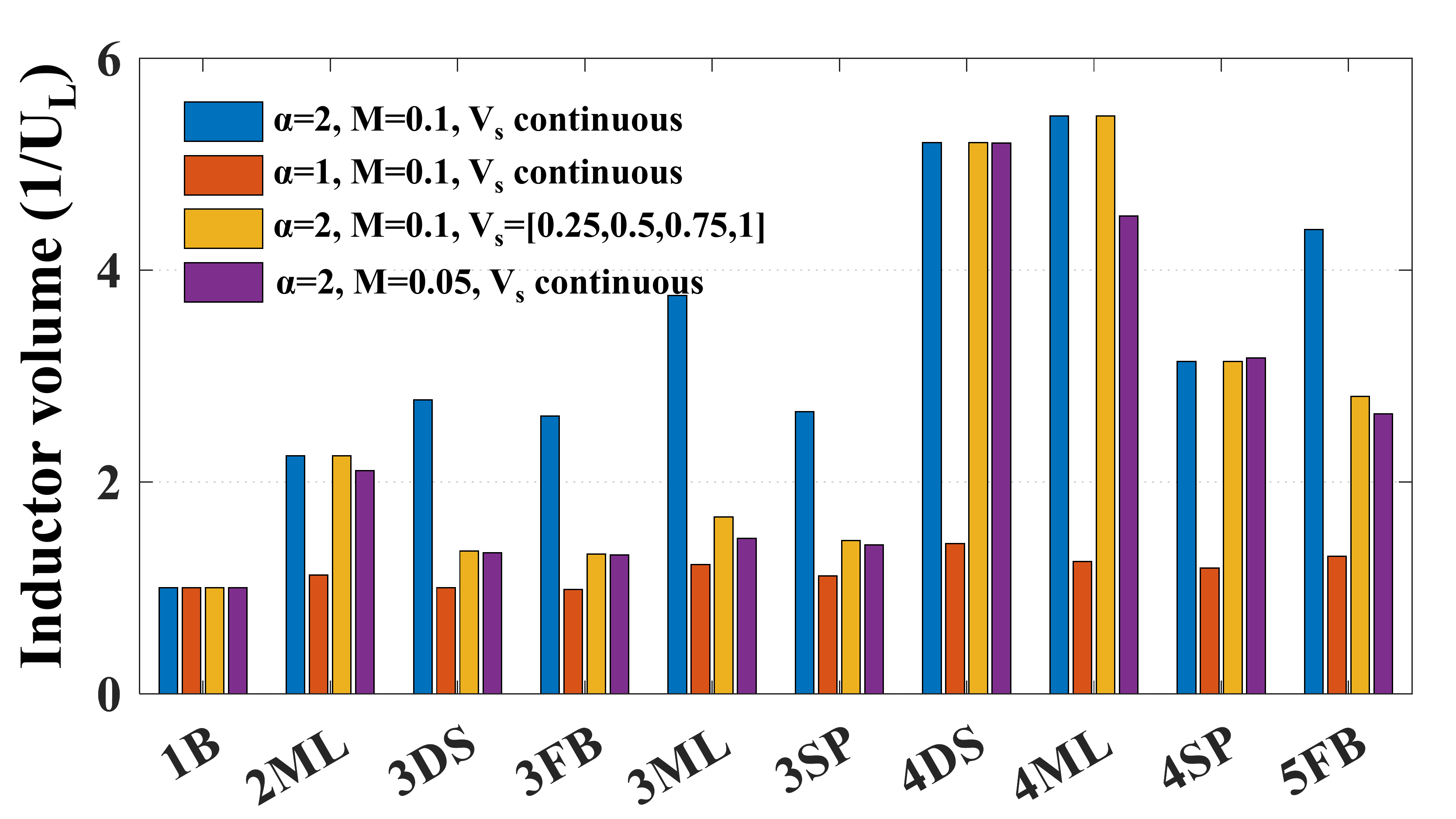}
    \caption{Input variables influence on inductor volume reduction ($\beta$=0).}
    \label{fig:5}
    \end{center}
\end{figure}

As noted in the introduction, our analysis has been limited to an ideal sizing process that assumes access to the best specific resistance for each voltage rating. However, off-the-shelf components or silicon integrated technology often do not offer such a wide range of voltage-rating options, which can limit the practical use of topologies with various values in the $V_S$ vector. By examining the number of different values in the $V_S$ vector, designers can easily evaluate the potential switch flavors required. For readers interested in quantifying this impact, we present in Fig.~\ref{fig:5} (yellow bars) the effect of quantizing $V_S$ (into 4 discrete values) instead of using the optimal value of $V_S$ (blue bars). While 4ML and 4DS remain the best options, the benefits of the two flying capacitors structures are greatly reduced.

The coefficients of the switch scaling law ($\alpha$ and $\beta$) also significantly affect the conclusion. For instance, when the switching penalty is lower (i.e., lower $\alpha$ and $\beta$), the advantages of HSCC diminish, resulting in different rankings, and in some cases, 1B can be the best option. In Fig.~\ref{fig:5} (blue and red bars), we compare the inductor volume gain for two different switching scaling parameters ($\alpha$). When the switching penalty is lower, the 4DS topology outperforms the 4ML topology.

Moreover, the reduction in inductor volume also varies with the voltage conversion ratio. Fig.~\ref{fig:5} (violet bars) illustrates the case where $M=0.05$. Here, 4DS outperforms the other topologies, including 4ML. Additionally, the two flying capacitors topologies show no improvement compared to 1B and are therefore not recommended.

With some practice, designers can show the effect of each topological parameter $(C, S, V_s, V_c, m, d, p, s)$ on their particular focus, and it can also help in introducing a new topology. We have described the method such that any new emerging topologies can be analyzed and incorporated into the comparison space. The question is, which topological parameters favor volume and open-loop bandwidth to achieve superior performance? The inspection of equations aims to reduce the values of $C, V_s, V_C, S, s,$ and $d$ as much as possible and to increase $p$. Unfortunately, some parameters are intrinsically linked, such as $V_C$ and $V_S$. More voltage blocked across the flying capacitor (higher $V_C$) reduces the voltage rating requirement for the switches (lower $V_S$).

It should be noted that practical converter designs have additional considerations beyond power transistors and passive sizing, such as the VCR range, the number of switches, gate drivers, additional drain capacitance loss, level shifters, capacitor charge balancing, voltage stress in starting phase, feedback control, PCB routing, and EMI. This may put the low $N$ topology in an attractive position. However, all factors with weighted coefficients, depending on the application, must be considered for a final choice.

\section{Conclusion}

In this paper, we present a comprehensive framework for comparing non-resonant hybrid switched-capacitor converters. By adjusting the switch area $A$ and switching frequency $F$, we establish a relationship between the active and passive performance of each topology to determine the achievable reduction in inductor volume and gain in bandwidth at the same efficiency and ripples as the conventional buck converter. Our discussion offers practical guidelines for designers to leverage this framework in their decision-making process, and can also be used to evaluate the performance of newly proposed topologies.

\appendix[Variables Definition]

Although each topological, design, and input variable is defined in the paper, we summarize them below for convenience.

The topology terminologies are:
\begin{itemize}
    \item 1B: 1-phase 2-level Buck \cite{erickson_fundamentals_2001} 
    \item NML: N:1 Flying-Capacitor Multi-Level \cite{lei_analysis_2013,lei_general_2015, rentmeister_924_2020,ye_design_2017}
    \item NSP: N:1 Series-Parallel \cite{pilawa-podgurski_merged_2008, lei_analysis_2013,lei_general_2015}
    \item NFB: N:1 Fibonacci Hybrid Converters \cite{yamauchi_055_2021,lei_general_2015}
    \item NDS: N:1 Hybrid Dickson Switched-Cap. Conv \cite{lei_soft-charging_2014, assem_hybrid_2020,lei_analysis_2013,lei_general_2015}
\end{itemize}

The \textit{topological} variables that describe the topology are:
\begin{itemize}
\item The current vector $C$ represents the RMS current flowing through each switch during the entire switching period $T$. Each element is normalized with respect to $I_O$. This notation is similar to the charge multiplier used in \cite{seeman_analysis_2008}.
\item The switching rate activity of each switch is given by each element of the vector $S$. The value of $N$ means the switch considered commutes $N$ times during $T$.
\item Each element of the vector $V_s$ represents the maximal voltage experienced by each switch during all states. All components are normalized with respect to $\widetilde{V_{in}}$.
\item The voltage blocked by each flying capacitor is described in the vector $V_c$, where each element is the normalized voltage referenced to $\widetilde{V_{in}}$. A unity value means the DC voltage across the capacitor is $\widetilde{V_{in}}$.
\item The scalar $m_{k}$ represents the normalized $V_{Lx}$ voltage by dividing the maximal $\widetilde{V_{Lx}}$ by $\widetilde{V_{in}}$.
\item The relationship between the desired voltage conversion ratio $M$ and the duty cycle $D$ is illustrated by the normalized duty-cycle $d_{k}=\frac{D}{M}$.
\item The normalized apparent switching frequency of $V_{Lx}$ is given by the scalar $p$. The value of $N$ means $V_{Lx}$ switches $N$ times during $T$.
\item The minimal normalized equivalent flying capacitor seen from the inductor from any state during $T$ is quantified using the scalar $s$. The value is normalized with respect to $C_F$.
\end{itemize}

The \textit{input} variables defining the design context are:
\begin{itemize}
\item The voltage conversion ratio ($M$).
\item The active device scaling law ($\alpha$ and $\beta$).
\end{itemize}

The \textit{design} variables resulting from the framework regarding 1B are:
\begin{itemize}
\item The relative switch area ($A$) required to obtain the same output resistance as 1B.
\item The relative switching frequency ($F$) required to obtain the same switching loss as 1B.
\item The relative inductor value ($L$) required to obtain the same inductor ripple as 1B.
\item The relative output capacitor value ($C_o$) required to obtain the same voltage ripple as 1B.
\item The relative flying capacitor value ($C_f$) required to obtain the same ratio between the switching frequency and the minimal self-resonance frequency formed by the combination of $C_F$ and $L$.
\end{itemize}

Each relative value $x_k$ of topology $k$ can be denormalized to $\widetilde{x}$ with respect to the absolute value of $x$ for topology 1B by:
\begin{equation}
\widetilde{x_k}=x_k\widetilde{x_{1B}}
\end{equation}

\bibliography{refs}

\end{document}